# Superconducting TiN Films Sputtered over a Large Range of Substrate DC Bias

H. M. Iftekhar Jaim, J. A. Aguilar, B. Sarabi, Y. J. Rosen, A. N. Ramanayaka, E. H. Lock, C. J. K. Richardson, and K. D. Osborn

*Abstract*— We have investigated properties of superconducting titanium nitride (TiN) films that were sputtered over a large range of RF-induced DC bias voltage applied to the substrate. Films grown with the largest bias voltages contained cubic TiN phases with a large fraction of the (200) crystalline growth orientation. These films also contained the smallest concentrations of oxygen impurities, resulting in stoichiometric TiN. Over the range of bias, variations of the stress from slightly tensile to highly compressive were measured and correlated to crystallinity of the (200) growth. The films exhibited highly uniform thickness and resistivity, and show the potential for yielding reproducible low-temperature devices. Finally, coplanar resonators fabricated with the films exhibited high kinetic inductance and quality factor, where the latter was obtained in part from temperature-dependent frequency shifts.

## I. Introduction

TITANIUM NITRIDE (TiN) thin films have recently demonstrated advantageous properties for superconducting quantum computing and astrophysics applications. The high quality factors of TiN coplanar resonators, on the order of $10^7$ [1], result in long lifetimes for quantum information storage circuits [1]-[3], and microwave kinetic inductance detectors (MKID) have potential for single photon sensitivity at low temperatures [4]. Furthermore, the high kinetic inductance of TiN have been used in microwave parametric amplifiers [5] and show potential for use in coherent phase slip qubits [6].

Even though there are advances in growth of this material, significant challenges remain for practical device applications. Film properties can be tuned by controlling the flow of $N_2$ [7], substrate temperature [8]-[9], chamber pressure, and distance between the target and the substrate [10], some of which were grown with a 100V DC bias [8]. However, some of the best TiN films exhibit non-uniform resistivity, $T_C$, and crystalline phases [10], and contain oxygen contaminants [11]. Oxygen is known to be incorporated after growth, such that aging effects are a possible concern. However, perhaps more serious, is the possibility that oxygen at the surface of the superconductor can induce dielectric loss or associated resonator phase noise, since oxides are known to contain two-level systems (TLS) [12]. As a result, it is desirable to have a superconducting TiN deposition technique which produces low-contaminant uniform films and which can be characterized in terms of its constituent phases.

In this paper we report on TiN sputtered films grown at different RF-induced DC bias voltages on the substrate (as a single varying parameter), where the DC biasing was explored up to 400V. Stress, resistivity, film composition, and crystalline phases were found to vary with the bias voltage. These conditions resulted in superconducting TiN films with high uniformity in resistivity and $T_C$ across the wafer, and with low impurity concentrations at the largest DC biases. Furthermore, resonators fabricated using films grown at high DC bias were characterized and revealed very high internal quality factors and high kinetic inductance fractions.

## II. Experimental Procedures

### A. TiN film deposition

TiN films were deposited by DC magnetron sputtering on high resistivity (>20,000 $\mu\Omega$-cm, float-zone) 3 inch silicon (100) substrates. Each silicon wafer was cleaned with 49 % HF for 30 seconds to remove the native oxide, and then inserted inside a load-lock within 5 minutes. The wafers were then transferred in-situ to the deposition chamber (base pressure $2\times10^{-8}$ Torr) within half an hour. The distance between the Ti target and the wafers was 20 cm.

During deposition, the substrate temperature was kept at 500 °C, while a constant pressure of 4.28 mTorr was maintained in the presence of Ar and $N_2$, flowed at 15 sccm and 10 sccm respectively. A power of 400 W was applied to a 3 inch Ti target for all sputtering. Pre-sputtering was performed for 1 minute without the RF-bias, and an additional 30 seconds with the RF power applied to the substrate prior to opening the substrate shutter. Different samples were fabricated by applying different RF power on the substrate. The power was chosen to maintain a particular DC bias voltage between 0 and 400 V. Once the

H. M. Iftekhar Jaim is with the Laboratory for Physical Sciences, College Park, MD 20740 USA, and the Department of Materials Science and Engineering, University of Maryland, College Park, MD 20742 USA, e-mail: jaim@umd.edu.

J. A. Aguilar is with the Department of Physics and Astronomy, Johns Hopkins University, Baltimore, MD 21218 USA, and is a National Physical Science Consortium Graduate Fellow, e-mail: jaguilar@pha.jhu.edu.

B. Sarabi, Y. J. Rosen, and A. N. Ramanayaka, are with the Laboratory for Physical Sciences, College Park, MD 20740 USA and the Department of Physics, University of Maryland, College Park, MD 20742 USA. e-mail: bahmans@lps.umd.edu, yrosen@lps.umd.edu, arunaniresh@lps.umd.edu.

E. H. Lock is with the Plasma Physics Division, Naval Research Laboratory, Washington, DC, e-mail: evgeniya.lock@nrl.navy.mil. This author acknowledges funding from the Naval Research Laboratory Base Program and NSI MIPR number # H98230201771.

C. J. K. Richardson is with the Laboratory for Physical Sciences, College Park, MD 20740 USA, email: richardson@lps.umd.edu.

K. D. Osborn is with the Laboratory for Physical Sciences, College Park, MD 20740 USA, and is the corresponding author. Phone: 301-935-6969; e-mail: osborn@lps.umd.edu.

deposition rate was determined for each voltage, the deposition time was adjusted to yield films with a nominal thickness of 62 nm.

*B.    Film characterization*

The crystal structure and phases of the deposited films were determined using X-ray diffraction (XRD). A θ-2θ scan was performed on the samples in the range 27°-60° (Fig. 1). All films showed a strong Si peak at 33°-34° from the substrate. For the 0 V DC bias film, XRD showed the presence of α-Ti, $TiO_2$ and TiN. Here TiN exhibited both (111) and (100) growth, where the latter was observed as the (002) plane and we adopt the common label as (200) growth. For the 110 V bias sample, weak peaks from both (111) and (200) TiN growth indicate a lack of dominant crystal orientation and could indicate an amorphous microstructure. As the bias was increased to 175 V and higher, the (111) peak disappeared and the (200) peak became dominant. For these samples, only a small peak of the tetragonal phase, $Ti_2N$, a normal conductor, was seen. Finally, at the highest bias measured (400 V), the (200) peak broadened and reduced in intensity. The background intensity of all samples was much higher than that of a reference bare silicon crystal wafer, and revealed some amorphous character to the TiN films.

The stress was measured on three sets of wafers containing the DC substrate biases, and the data showed reproducible stress for each bias. Fig. 2 shows the TiN film stress against the left axis and the (200) peak height normalized to the background value from the XRD measurements against the right axis, for the different substrate biases on the bottom axis. The stress varied from slightly tensile to highly compressive across the range of bias conditions, and shows a correlation with the changes in (200) growth fraction [13].

Sputter elemental depth profiles were obtained in a K-Alpha X-ray photoelectron spectroscopy (XPS) system using a beam of high-energy argon ions. The profile of oxygen is shown as a function of sputtering time in Fig. 3a. and elemental composition of TiN films at a depth of approximately 30 nm is shown in (Table I). The TiN films showed almost no carbon contamination in their bulk, and at substrate biases above 110 V no oxygen was detected either. At the lower substrate biases, the oxygen content appears to decrease as a function of the sputtering time, and this might be due to reduction of oxygen contamination permeating through the grain boundaries due to a denser film. Furthermore, the oxygen content and resistivity (Fig. 3b) were correlated. The highest resistivity, 436 μΩ-cm, was observed in the 0 V bias film, which had 7 % oxygen at 30 nm below the surface. As substrate bias increased, a gradual reduction of oxygen content and resistivity was observed. When oxygen concentration reached 0 % the resistivity dropped to approximately 28 μΩ-cm, slightly above the single crystalline TiN value of 18 μΩ-cm [14]. Similar trends were observed in previous work [15].

The uniformity of the film thickness, and complex index of refraction was characterized using broadband spectrophotometry. The films exhibited highly uniform thicknesses with relative standard deviations below 2 % over the wafer. The index of refraction, n, had relative standard deviations below 5 %, and the extinction coefficient, k, less than 2 %.

Room-temperature resistivity was measured using a 4-point probe with a contact spacing of 0.050 inch on unpatterned films, deposited on 3 inch diameter wafers. Resistivity was probed using a 0.5 inch grid across the entire wafers, which resulted in 21-25 data points per wafer. Reproducibility of the resistivity was analyzed using three films grown at 110V DC bias and two others grown at 250 V DC bias. Resistivity was found to vary by less than 2 % within any single film, and within 2 % between different films deposited at the same bias voltage. Superconducting critical temperature measurements were also performed. As the substrate bias was increased, the $T_C$, measured using a 90% resistance criteria, showed a slight increase (Table I), but the film $T_C$ was generally in the range of 4 K < $T_C$ < 4.5 K. The $T_C$ also showed deviations of less than 2% in comparing measurements at the center and edge of the 3 inch wafers. This can be compared to a different growth technique which yielded a $T_C$ variation of ±15% across the wafer [23].

*C.    Resonator Fabrication*

Quarter-wave coplanar waveguide resonators were fabricated to investigate the TLS-induced surface loss and kinetic inductance of the TiN films. After TiN deposition and lithography, the film was etched using a chlorine-based recipe ($BCl_3$ 5:1 $Cl_2$) in an inductively coupled plasma etcher. To remove the leftover Cl content, samples were then cleaned with phosphoric acid. After this process, energy-dispersive x-ray spectroscopy showed no trace of Cl.

To avoid impedance mismatches to a 50 ohm off-chip coplanar waveguide [16], the center and the edges of the ground plane of the on-chip coplanar transmission line were made of aluminum (see inset of Fig. 4). For this, 120 nm of aluminum was deposited by e-beam evaporation through a patterned bilayer of resist which was developed and cleaned in Fujifilm OPD4262 and DI water, respectively, prior to depositing on a region with TiN. SEM imaging and profilometry confirmed that no fencing occurred on the edges. The TiN in the resonators were in contact with Fujifilm OIR resist during the TiN etching step and the Microchem LOR resist, as the bottom layer of a bilayer, during the aluminum liftoff step. These resists were removed with acetone and Microchem Remover PG, respectively.

*D.    Low-temperature microwave measurements*

The TiN film thickness in all fabricated samples was approximately 62 nm, which is much smaller than the London penetration depth previously measured for TiN (250 nm – 700 nm) [1]. Resonator 1 and 2 were fabricated with center conductor widths of 3 μm and 20 μm, and gaps between center conductor and ground plane of 3 μm and 15 μm, respectively. Resonators were measured in a dilution refrigerator in the temperature range of 0.03 K < $T$ < 1.3 K . The designed frequencies for the two resonators using a zero penetration depth, $f_{\lambda=0}$, are equal to 7.5 and 6.9 GHz, respectively. To date, successfully analyzed data only exists for the 250 and 400 V DC bias samples, where the inductive shift is generally smaller than the lowest bias samples due to the lower resistivity. Fig. 4 shows the fractional frequency shift from a 400 V bias film as a function of the temperature for two resonators, measured at the same power and at less than or equal to 50 photons stored.

The fractional frequency shift for both the resonators showed a non-monotonic variation verses the temperature [17]. At low temperatures, $T < 0.25$ K, the fractional frequency shift was observed to increase with temperature, which is believed to be caused by TLSs. Therefore, the low-temperature frequency shift was fit to the logarithmic term from Ref. [18], which is known to dominate at the lowest temperatures. From this fit, shown in Fig. 4b, we extracted TLS internal quality factors of $Q_{i\,TLS} = 3 \times 10^5$ and $Q_{i\,TLS} = 1 \times 10^6$ for resonators 1 and 2, respectively. This qualitatively agrees with expectations because the fractional filling volume from TLS-containing surfaces is smaller in resonator 2. The total internal quality factors extracted from analysis [19] of the 250 V bias resonator had an internal quality of approximately $3 \times 10^5$ at single photon energy, and the 400V bias sample was also very high but they were too inconsistent to merit a comparison to the TLS internal quality factors discussed above. The 400 V and 250 V bias films, which have a very high Q, also have a relatively high concentration of the (200) growth orientation relative to the (111) orientation.

The kinetic inductance, $L_k$, can be related to the total inductance $L_{tot}$ by the kinetic inductance fraction $\alpha = L_k/L_{tot}$. To extract the this quantity, the entire temperature range, 0.05 K – 1.3 K, was fit to a model combining the lowest-temperature TLS induced effects and shift $\frac{\delta f(T)}{f(0)} = -\frac{\alpha_{MB}}{2}\sqrt{\frac{\pi\Delta_0}{2k_B T}}e^{-\frac{\Delta_0}{k_B T}}$, from the Mattis-Bardeen formula (e.g. eq. 11 of [20]), while keeping the TLS fit (described above) unchanged. This second fit yielded the kinetic inductance fraction of $\alpha_{MB} = 47\%$ and 23% for resonators 1 and 2, respectively. The same fit yielded $\Delta/k_B T_C = 2.0$ ($T_C$ = 4.32 K). From the measured resonator frequency $f_{meas}$ compared to the single simulated value, we calculate $\alpha_{approx} = 1 - (f_{meas}/f_{\lambda=0})^2$ equal to 70% and 29% for resonator 1 and 2, respectively. This quantity is expected to approximate $\alpha$ only when the geometric inductance, $L_{tot} - L_k$, is not changed significantly by the finite penetration depth [18]. Among these resonators, resonator 2 has a significantly smaller $\alpha$ because of a significantly the larger center conductor width compared to the 2D screening length [21]. It also has a smaller geometric inductance such that one expects $\alpha_{approx}$ to be a better approximation of $\alpha_{MB}$ than for resonator 1, which is in agreement with the data.

### III. DISCUSSION

The orientations of the TiN were dependent upon the bias voltages. At low bias voltages the presence of a (111) growth plane was measured. This can be due to the lower strain energy from this phase [22]. However, as the bias was raised, a decrease in the sharpness of the (200) XRD peak was seen up to 250 V. This phase is known to grow from a SiNx seed layer [1].

The elemental composition of TiN films at 30 nm showed considerable reduction of oxygen content with biasing above 175 V and these voltages also produced stoichiometric TiN films. This is probably caused by the production of denser films with fewer voids and open structures for oxygen diffusion [15], while all films had low C content. The presence of oxygen also seemed to correlate with film resistivity. However, it is important to note that, in contrast to previous studies [10], our films were highly uniform in thickness, resistivity, optical properties, and $T_C$ across the film. The resistivity was consistent between depositions, and was also found to be nearly unchanged over several months after the deposition.

The superconducting temperature was fairly constant for all the films. For the 0 V DC bias film, the presence of excess Ti ($T_C$ = 0.4 K) with TiN (pure TiN, $T_C$ = 5.6 K) [23] might have caused the reduction in $T_C$ to 4.12 K. A slight increase in the $T_C$ was seen as substrate bias was increased, but they remained below 5 K. Low-temperature measurements of resonators showed large values of the kinetic inductance fraction $\alpha_{MB}$ for quarter-wave resonators. Large kinetic inductances were also found by comparing the resonator frequency to the simulated frequency for a zero penetration depth film.

### IV. CONCLUSION

We studied a method of controlling growth in superconducting TiN films using a large range of RF-induced DC biases on the substrate. With this method, we were able to reduce the oxygen content within the film to negligible levels, while the carbon content was generally negligible in the bulk. We were also able to achieve a wide range of resistivity while maintaining TiN film property uniformity. The variable stress was related to the crystalline phases. $T_C$ only exhibited a small bias dependence and this could be related to the stoichiometry as well as the crystalline phases. Finally, TiN quarter-wave resonators were fabricated and showed high internal quality factors with a TLS internal quality factor extracted using a temperature-dependent frequency shift at the lowest temperatures. A larger temperature range of the shift revealed the large kinetic inductance in the films. By studying a large range of the substrate DC bias, induced by RF power, we found superconducting TiN films with superior purity and a 2 % uniformity could be produced over a range of different crystalline phases. As a result this growth technique should be useful for many device applications where uniformity and reproducibility of kinetic inductance and quality factor are important.

### V. ACKNOWLEDGEMENTS

The authors acknowledge helpful discussions with M. J. A. Stoutimore, M. Sandberg, and D. P. Pappas.

TABLE I: THE DEPOSITION RATE, ELEMENTAL CONCENTRATIONS, $T_C$ AND RESISTIVITY OF THE DIFFERENT FILMS FOR DIFFERENT BIAS VOLTAGES. THE XPS ELEMENTAL COMPOSITION WAS OBTAINED AT 30 NM BELOW THE TiN FILM SURFACES.

| Bias Voltage (V) | | 0 | 110 | 175 | 250 | 325 | 400 |
|---|---|---|---|---|---|---|---|
| Deposition Rate (nm/min) | | 2.15 | 1.82 | 1.52 | 1.57 | 1.51 | 1.57 |
| XPS (at. %) | Ti | 47.9 | 47.8 | 49.7 | 49.3 | 50 | 50 |
| | N | 44.7 | 47.5 | 50.1 | 50.7 | 50 | 50 |
| | O | 7.2 | 3.7 | 0.1 | 0 | 0 | 0 |
| | C | 0.2 | 0 | 0 | 0 | 0 | 0 |
| $T_C$ (K) | | 4.12 | 4.37 | 4.36 | 4.45 | 4.58 | 4.35 |
| R (μΩ-cm) | | 436 | 96 | 64 | 29 | 22 | 32 |

## VI.　Image Files

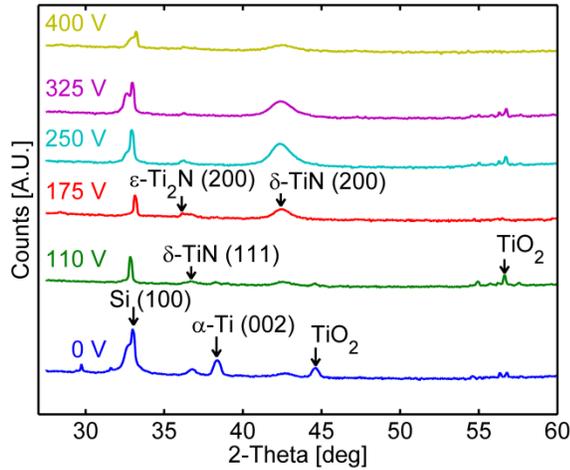

Fig. 1. XRD of the TiN films deposited at RF-induced DC bias of 0 V, 110 V, 175 V, 250 V, 325 V and 400 V. Labels are added to indicate the positions of relevant phases for Ti-N growth.

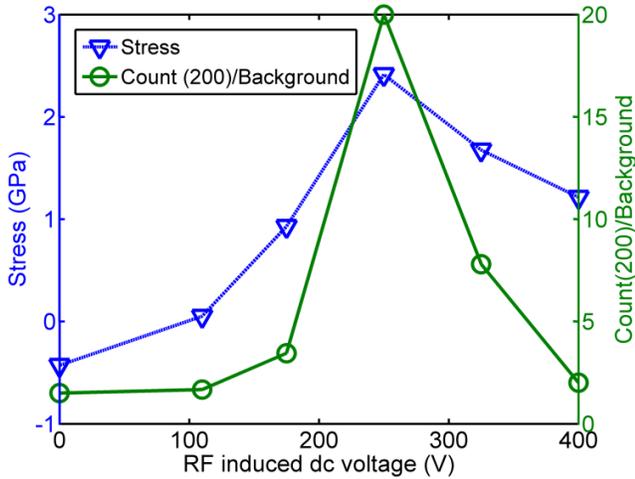

Fig. 2. Stress (compressive is positive) and relative XRD intensity along (200) growth of the TiN films as a function of the RF induced DC voltage applied to the substrate.

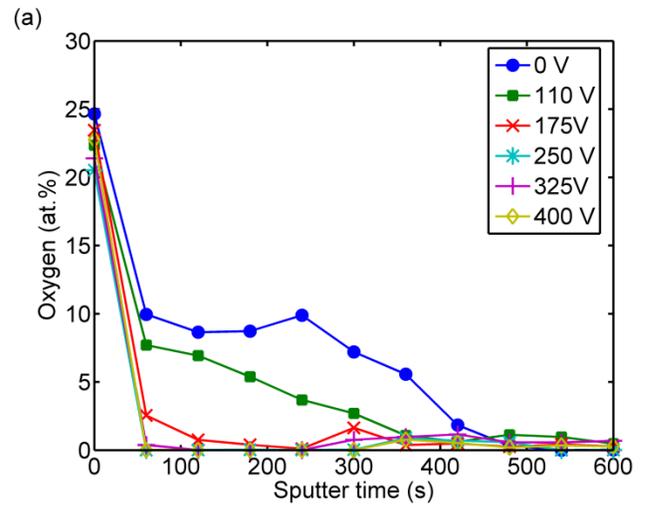

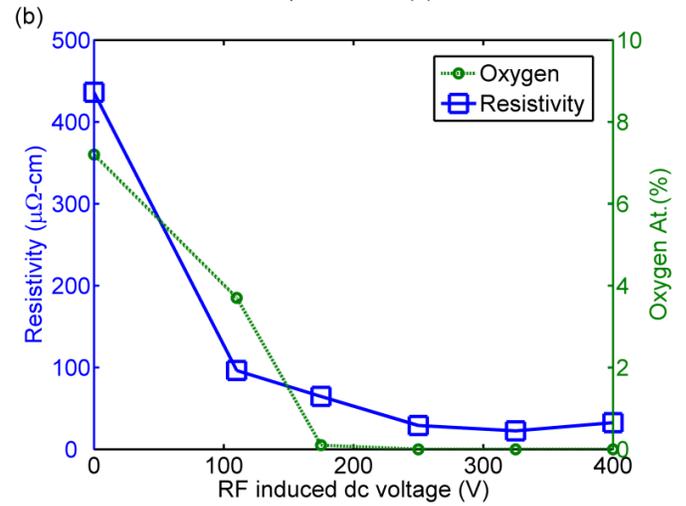

Fig. 3. (a) Oxygen contents of the TiN films analyzed by XPS as a function of Ar etch time. (b) The oxygen content 30 nm below the film surface and resistivity as a function of the RF-induced DC bias voltage applied to the substrate.

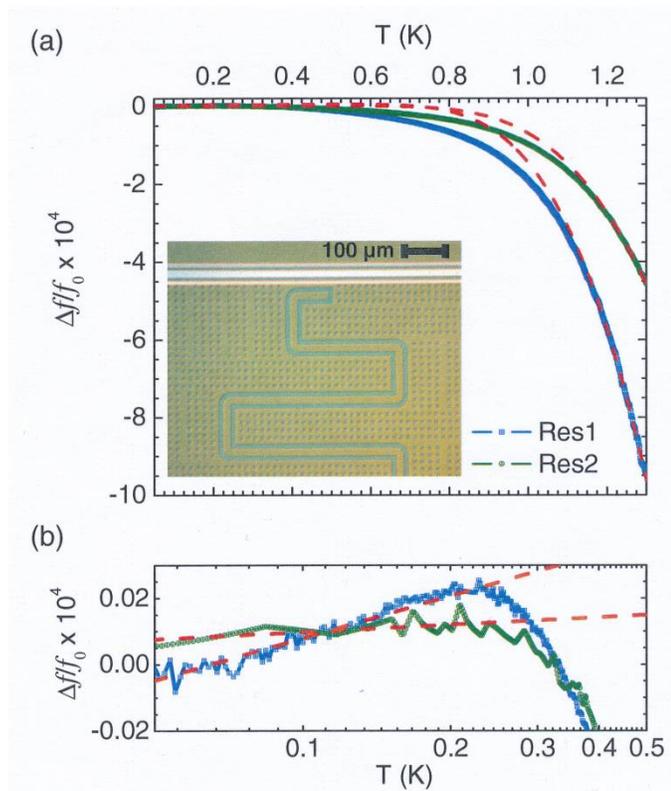

Fig. 4. (a) The measured fractional frequency shift for two resonators as a function of the temperature for the 400 V sample. The red dashed lines are fits to a high and low temperature theory. Inset shows an optical image of a quarter-wave resonator. TiN, Al, and the Si substrate appear in green, white and gray, respectively. (b) An enlarged view of the low temperature data from panel (a) showing the non-monotonic behavior of the fractional frequency shift at low temperatures. The fits to the TLS induced frequency shift are also shown (red dashed lines).